\documentstyle[12pt,prd,aps,multicol,psfig,epsf]{revtex}
\begin{document}
\title{
\hfill hep-ph/9903435\\
Sneutrino Vacuum Expectation Values and Neutrino Anomalies Through Trilinear
R-parity Violation}
\author{ Anjan S.~Joshipura 
 and  Sudhir K.~Vempati}
  \address{ Theoretical Physics Group, Physical Research Laboratory,
   Navarangpura, Ahmedabad, 380 009, India.}
\def\ap#1#2#3{           {\it Ann. Phys. (NY) }{\bf #1} (19#2) #3}
\def\arnps#1#2#3{        {\it Ann. Rev. Nucl. Part. Sci. }{\bf #1} (19#2) #3}
\def\cnpp#1#2#3{        {\it Comm. Nucl. Part. Phys. }{\bf #1} (19#2) #3}
\def\apj#1#2#3{          {\it Astrophys. J. }{\bf #1} (19#2) #3}
\def\asr#1#2#3{          {\it Astrophys. Space Rev. }{\bf #1} (19#2) #3}
\def\ass#1#2#3{          {\it Astrophys. Space Sci. }{\bf #1} (19#2) #3}

\def\apjl#1#2#3{         {\it Astrophys. J. Lett. }{\bf #1} (19#2) #3}
\def\ass#1#2#3{          {\it Astrophys. Space Sci. }{\bf #1} (19#2) #3}
\def\jel#1#2#3{         {\it Journal Europhys. Lett. }{\bf #1} (19#2) #3}

\def\ib#1#2#3{           {\it ibid. }{\bf #1} (19#2) #3}
\def\nat#1#2#3{          {\it Nature }{\bf #1} (19#2) #3}
\def\nps#1#2#3{          {\it Nucl. Phys. B (Proc. Suppl.) }
                         {\bf #1} (19#2) #3} 
\def\np#1#2#3{           {\it Nucl. Phys. }{\bf #1} (19#2) #3}
\def\pl#1#2#3{           {\it Phys. Lett. }{\bf #1} (19#2) #3}
\def\pr#1#2#3{           {\it Phys. Rev. }{\bf #1} (19#2) #3}
\def\prep#1#2#3{         {\it Phys. Rep. }{\bf #1} (19#2) #3}
\def\prl#1#2#3{          {\it Phys. Rev. Lett. }{\bf #1} (19#2) #3}
\def\pw#1#2#3{          {\it Particle World }{\bf #1} (19#2) #3}
\def\ptp#1#2#3{          {\it Prog. Theor. Phys. }{\bf #1} (19#2) #3}
\def\jppnp#1#2#3{         {\it J. Prog. Part. Nucl. Phys. }{\bf #1} (19#2) #3}

\def\rpp#1#2#3{         {\it Rep. on Prog. in Phys. }{\bf #1} (19#2) #3}
\def\ptps#1#2#3{         {\it Prog. Theor. Phys. Suppl. }{\bf #1} (19#2) #3}
\def\rmp#1#2#3{          {\it Rev. Mod. Phys. }{\bf #1} (19#2) #3}
\def\zp#1#2#3{           {\it Zeit. fur Physik }{\bf #1} (19#2) #3}
\def\fp#1#2#3{           {\it Fortschr. Phys. }{\bf #1} (19#2) #3}
\def\Zp#1#2#3{           {\it Z. Physik }{\bf #1} (19#2) #3}
\def\Sci#1#2#3{          {\it Science }{\bf #1} (19#2) #3}
\def\n.c.#1#2#3{         {\it Nuovo Cim. }{\bf #1} (19#2) #3}
\def\r.n.c.#1#2#3{       {\it Riv. del Nuovo Cim. }{\bf #1} (19#2) #3}
\def\sjnp#1#2#3{         {\it Sov. J. Nucl. Phys. }{\bf #1} (19#2) #3}
\def\yf#1#2#3{           {\it Yad. Fiz. }{\bf #1} (19#2) #3}
\def\zetf#1#2#3{         {\it Z. Eksp. Teor. Fiz. }{\bf #1} (19#2) #3}
\def\zetfpr#1#2#3{         {\it Z. Eksp. Teor. Fiz. Pisma. Red. }{\bf #1} (19#2) #3}
\def\jetp#1#2#3{         {\it JETP }{\bf #1} (19#2) #3}
\def\mpl#1#2#3{          {\it Mod. Phys. Lett. }{\bf #1} (19#2) #3}
\def\ufn#1#2#3{          {\it Usp. Fiz. Naut. }{\bf #1} (19#2) #3}
\def\sp#1#2#3{           {\it Sov. Phys.-Usp.}{\bf #1} (19#2) #3}
\def\ppnp#1#2#3{           {\it Prog. Part. Nucl. Phys. }{\bf #1} (19#2) #3}
\def\cnpp#1#2#3{           {\it Comm. Nucl. Part. Phys. }{\bf #1} (19#2) #3}
\def\ijmp#1#2#3{           {\it Int. J. Mod. Phys. }{\bf #1} (19#2) #3}
\def\ic#1#2#3{           {\it Investigaci\'on y Ciencia }{\bf #1} (19#2) #3}
\def\tp{these proceedings}
\def\pc{private communication}
\def\ip{in preparation}
\relax

\newcommand{\Tev}{\,{\rm Tev}}
\newcommand{\GeV}{\,{\rm GeV}}
\newcommand{\MeV}{\,{\rm MeV}}
\newcommand{\keV}{\,{\rm keV}}
\newcommand{\eV}{\,{\rm eV}}
\newcommand{\Tr}{{\rm Tr}\!}
\renewcommand{\arraystretch}{1.2}
\newcommand{\beq}{\begin{equation}}
\newcommand{\eeq}{\end{equation}}
\newcommand{\beqa}{\begin{eqnarray}}
\newcommand{\eeqa}{\end{eqnarray}}
\newcommand{\ba}{\begin{array}}
\newcommand{\ea}{\end{array}}
\newcommand{\bmat}{\left(\ba}
\newcommand{\emat}{\ea\right)}
\newcommand{\refs}[1]{(\ref{#1})}
\newcommand{\ler}{\stackrel{\scriptstyle <}{\scriptstyle\sim}}
\newcommand{\ger}{\stackrel{\scriptstyle >}{\scriptstyle\sim}}
\newcommand{\lag}{\langle}
\newcommand{\rag}{\rangle}
\newcommand{\ns}{\normalsize}
\newcommand{\cm}{{\cal M}}
\newcommand{\gr}{m_{3/2}}
\newcommand{\p}{\partial}

\def\rp{ $R_P$} 
\def\321{$SU(3)\times SU(2)\times U(1)$}
\def\tl{{\tilde{l}}}
\def\tL{{\tilde{L}}}
\def\bd{{\overline{d}}}
\def\tL{{\tilde{L}}}
\def\a{\alpha}
\def\b{\beta}
\def\g{\gamma}
\def\c{\chi}
\def\d{\delta}
\def\D{\Delta}
\def\db{{\overline{\delta}}}
\def\Db{{\overline{\Delta}}}
\def\e{\epsilon}
\def\k{\kappa}
\def\l{\lambda}
\def\n{\nu}
\def\m{\mu}
\def\nt{{\tilde{\nu}}}
\def\om{\omega}
\def\p{\phi}
\def\P{\Phi}
\def\x{\xi}
\def\r{\rho}
\def\s{\sigma}
\def\t{\tau}
\def\th{\theta}
\def\ne{\nu_e}
\def\nm{\nu_{\mu}}
\def\rp{$R_P$}
\def\mp{$M_P$}     
\renewcommand{\Huge}{\Large}
\renewcommand{\LARGE}{\Large}
\renewcommand{\Large}{\large}
\maketitle

\begin{center} Abstract \end{center}

\begin{abstract}
Neutrino mass spectrum is reanalyzed in supersymmetric models
with explicit trilinear $R$ violation. Models in this category are argued
to provide simultaneous solution to the solar and atmospheric neutrino
anomalies. It is shown specifically that
large mixing and hierarchical masses needed for the vacuum solution of 
neutrino anomalies arise  naturally in these models without 
requiring any  additional symmetries or hierarchies among the 
trilinear couplings.
\end{abstract}

\noindent {\bf 1.~Introduction:}

\noindent The hypothesis of neutrino oscillations has gained acceptance 
after careful observation of the atmospheric muon neutrino deficit
at the Superkamioka \cite{sk}.  If neutrinos do oscillate then both the solar
 and the atmospheric neutrino deficits can be 
simultaneously understood in terms of  vacuum oscillations among the three
known neutrinos. This however requires the presence of two large
mixing \cite{bimax}
angles among three neutrino states.
Many different theoretical models \cite{models}  have been proposed 
in this context. Supersymmetry (SUSY) provides a framework where both the
largeness of mixing and hierarchy in masses can 
be naturally understood.
 
Supersymmetric extension of the standard model contains the following
lepton number violating terms:
\beq \label{wl}
W_{\not L} = \l'_{ijk} L_i Q_j D_k^c + \l_{ijk} L_i L_j E_k^c + \e_i L_i H_2
\;\; .\eeq
These naturally lead to neutrino masses \cite{hs}. The neutrino 
spectrum in this
model has been extensively studied \cite{rest,gauge,drees,chun,arc,bs}
 in the literature in recent times.
It has been shown \cite{rest,gauge} that bilinear SUSY violating
 interactions provide very
economical framework which can simultaneously accommodate hierarchical
masses and large mixing. In contrast, the neutrino spectrum implied by
the trilinear interactions $\lambda,\lambda'$ may appear arbitrary a
priori due to very large number  of such couplings. It was emphasized 
by Drees et al. \cite{drees} that
this is not the case and neutrino spectrum could be quite predictive
even in models with trilinear couplings. This can be understood
  from eq.(\ref{wl}) which in the absence of bilinear terms is
 invariant under a
 global U(1) symmetry with U(1) charges 1, -1, -2 for the fields
 $L, D^c, E^c$ respectively. This symmetry would thus prevent generation 
of neutrino masses if it was not broken by the down quark and charged
 lepton Yukawa couplings $h^D_k$ and $h_k^E$ respectively.
This means that the neutrino masses generated by the
trilinear interactions in eq.(\ref{wl}) are always
accompanied by the above Yukawa couplings and hierarchy in the latter 
gets
decoded into the neutrino masses if all the trilinear couplings are
assumed to be similar in magnitude. In the limit
of keeping only the b quark Yukawa coupling, one combination
of neutrino fields namely ~ $\lambda'_{i33}\nu_i$,  obtains a mass and
this combination would contain large mixing of all the states if
$\lambda'_{i33}$ are comparable for different ~$i$. The other mass
would arise when the strange quark Yukawa coupling is turned on and one
thus naturally gets \cite{drees},
\beq \label{heir}
{m_{\nu_{\mu}}\over m_{\nu_{\tau}}}\sim {m_s\over m_b} \;\; ,
\eeq
reproducing the hierarchy needed to understand the solar and atmospheric
neutrino anomalies simultaneously. The natural expectation of this
scenario is large mixing among all three neutrinos and this is not
favored by the more likely small angle MSW \cite{msw} solution for the solar
neutrino deficit. This led to imposition of ad-hoc discrete symmetries in
\cite{drees} 
to prevent unwanted trilinear couplings reducing the attractiveness of the
scenario.

It is clear from the forgoing discussion that more natural possibility
with  the trilinear couplings of similar magnitudes is to have large
mixing
among all the neutrinos. This however then favors the vacuum solution
to the solar neutrino problem in which case the hierarchy among neutrino
masses is required to be
stronger than displayed in eq.(\ref{heir}). 
A careful analysis of the neutrino spectrum
reveals that under the standard assumptions, the neutrino mass hierarchy
resulting in models with only
trilinear $R$ violating couplings at a high scale is indeed stronger than
the one in eq.(\ref{heir}).
 It can be strong enough to get the vacuum solution for the solar
neutrino problem. This coupled with large mixing among neutrinos
alleviates any need to postulate  discrete symmetry as in
 \cite{drees} and makes the
trilinear lepton number violation an attractive means to understand neutrino
anomalies.

The key feature leading to a different conclusion compared to \cite{drees}
is the observation that the presence of trilinear interactions in
the original superpotential  at a high scale,
induces \cite{sneutrino} terms linear in the sneutrino fields
in the effective potential at the weak scale. 
 These sneutrino fields then obtain vacuum expectation value (vev) and cause
neutrino-neutralino mixing. Neutrino mass generated through this
mixing dominates over the loop mass considered in \cite{drees}
 and in  other works \cite{arc,other}. This
alters the neutrino mass hierarchy compared to eq.(\ref{heir}). We 
discuss these
issues quantitatively in the following.\\
{\bf 2. Sneutrino Vevs and Neutrino Masses }\\
For definiteness, we shall concentrate on the trilinear interactions
containing $\lambda'$ couplings and comment on the inclusion of the
$\lambda$ couplings latter on. The presence of non-zero $\lambda'_{ijk}$
is known to induce two separate contributions to the neutrino masses
and we discuss them in turn.\\
{\bf A. Tree level mass}\\
We adopt the conventional supergravity framework \cite{nilles}
 according to which the
structure of the superpotential dictates the structure of the soft SUSY
breaking terms. Thus, with only trilinear L-violating interactions, the
soft terms do not contain bilinear terms at a high scale. They are 
nevertheless generated at the weak scale \cite{sneutrino} and should be
retained in the scalar potential at this scale:
\beqa \label{soft}
V_{soft}&=& m_{\tilde{\n}_i}^2 \mid \tilde{\n}_i \mid ^2 + m_{H_1}^2 
\mid H_1^0 \mid^2 + m_{H_2}^2 \mid H_2^0 \mid^2  + \left[ m_{\n_i H_1}^2 
\tilde{\n}^{\star}_i H^0_1 \right. \nonumber \\
& & \left. - \m\;B_\m H_1^0 H_2^0 - B_{\e_i} \tilde{\n}_{i} H_2^0  + h.c
\right]
+ {1 \over 8} (g_1^2 + g_2 ^2) (\mid H_1^0 \mid  ^2 -
 \mid H_2^0 \mid ^2)^2 + .... \;\;\; .
\eeqa
Where, we have retained only neutral fields and used standard notation 
with $B_{\e_i}$ and $m_{\n_i H_1}^2$  representing the bilinear 
lepton number violating \cite{f1} soft terms.  The weak scale value of the soft 
parameters is determined by the following \cite{sneutrino} renormalization
group equations
(RGE):
\beqa \label{rge}
{d B_{\e_i} \over dt} & = & B_{\e_i} \left (- {1 \over 2} Y^{\tau} - {3 \over 2}
Y^t + {3 \over 2} \tilde{\a_2} + {3 \over 10} \tilde{\a_1} \right ) - 
{3 \over 16 \pi^2}~~ \m~~ h^D_{k}~~ \l'_{ikk}
 \left( {1 \over 2} B_\m + A^{\l'}_{ikk} \right ), \nonumber\\
{d m^2_{\n_i H_1} \over dt} & = & m^2_{\n_i H_1} \left( - 2 Y^{\tau}
- {3 \over 2} Y^b \right) - \left(3 \over 32 \pi^2 \right)~~ h^D_{k} ~~
\l'_{ikk} \left( m_{H_1}^2 + m_{L_i}^2 \right.\\ \nonumber
&~~& \left. +~~  2 ~~{m^2}_{kk}^{Q} + 2 ~~A^{\l'}_{ikk} 
A^D_{kk} + 2~~ {m^2}^{D^c}_{kk} \right),
\eeqa
and the standard RGE \cite{nilles} ~for the parameters on the RHS.
 Since we allow
only  trilinear interactions in $W_{\not L}$, $m_{\n_i H_1}^2 = B_{\e_i} = 0 $
 at high scale.
As seen from the above equations, the presence of non-zero $\l'_{ikk}$ however
generate non-zero values for $m_{\n_i H_1}^2$ and $B_{\e_i}$. It is then 
convenient to parameterize them as,
\beqa \label{sol}
B_{\e_i}&=& \l'_{ipp} h^D_{p} \k_{ip}, \nonumber \\
m_{\n_i H_1}^2&=&\l'_{ipp} h^D_{p} \k'_{ip}\;\;\;.
\eeqa
Here, $p$ is summed over generations. The parameters $\k$ and $\k'$ 
represent the 
running of  the parameters present in the RGE's from the GUT scale 
to the weak scale.

The above soft potential would now give rise to sneutrino vevs,
\beq
\label{omega}
< \tilde{\n_i} >~~   = {B_{\e_i} v_2 - m^2_{\n_i H_1} v_1 \over m_{L_i}^2 
+ {1 \over 2} m_Z^2~ cos 2 \beta }\;\;\;.
\eeq
The sneutrino vevs so generated will now mix the neutrinos with the neutralinos
 thus giving rise to a tree level neutrino mass matrix \cite{asjmarek}~:
\beqa
\label{mnot}
{\cal M}^0_{ij}&=&{\m (cg^2 + g'^2) ~ < \tilde{\n_i} >~ < \tilde{\n_j} > 
\over 2 ( -c \m M_2 + 2 M_w^2 c_\b s_\b (c + tan \theta_w^2 ))} \;\; .
\eeqa
{\bf B. Loop Level Mass }\\
The trilinear couplings in the superpotential would also give rise to
a loop induced neutrino mass with the down squark and antisquark pairs
being exchanged in the loops along with their ordinary partners 
\cite{hs,babu}.
This mass can be written as,
\beq
\label{mloop}
{\cal M}^{l}_{ij} = {3 \over 16 \pi^2 } \l'_{ilk} \l'_{jkl} ~ v_1~ h^D_{k} ~sin 
\phi_l\; cos \phi_l~ ln {M_{2l}^2 \over M_{1l}^2}\;\;\;. 
\eeq
In the above, $sin\phi_l\; cos\phi_l$ determines the mixing of the
 squark-antisquark pairs and $M_{1l}^2$ and $M_{2l}^2$ represent
the eigenvalues of the standard 2$\times$ 2 mass matrix of the 
down squark system \cite{nilles}. The indices $l$ and $k$ are
 summed over. The $v_1$ stands for the vev of the Higgs field $H^0_1$.
The mixing $sin\phi_l\; cos\phi_l$ is proportional to $h^D_l$ and thus 
one can write the loop mass as,
\beq 
\label{mloop1}
{\cal M}^l_{ij} = \l'_{ilk}~ \l'_{jkl}~ h^D_{k}~ h^D_{l}~ m_{loop}\;\; .
\eeq
Explicitly,
\beq
\label{expmloop}
m_{loop} \equiv {3 ~v_1 \over 16 \pi^2}~~ {sin \phi_{l}~ cos \phi_{l} 
\over h^D_l} ~~ln {M_{2l}^2 \over M_{1l}^2} \sim {3 ~v_1^2 \over 16 \pi^2}~~
 {1 \over M_{SUSY}},
\eeq
\noindent
with $M_{SUSY} \sim 1$ TeV referring to the typical scale of 
SUSY breaking. Note that
$m_{loop}$ defined above is independent of the R violating couplings and is 
solely determined by the  parameters of the minimal supersymmetric
standard model (MSSM).

As evident from eqs.(\ref{mnot},\ref{mloop}), the tree as well as loop
induced masses have very similar structure. Both involve down-quark 
Yukawa couplings for the reason explained in the introduction. The tree
level contribution involves diagonal couplings $\l'_{ikk}$ while the 
${\cal M}^l$ contains off-diagonal $\l'_{ikl}$ as well \cite{f2}.  
  If all $\l'_{ikl}$ are 
assumed to be of similar magnitude then the tree level mass is seen to
dominate over the loop mass as we will discuss in the next section.\\
{\bf 3. Neutrino Masses and Mixing}\\
We now make a simplifying approximation which allows us to discuss
neutrino masses and mixing analytically. It is seen
from the RG eqs.(\ref{rge}) that the parameters $\k_{ik}$,
$\k'_{ik}$, defined in 
eq. (\ref{sol}) are independent of generation structure in the limit in
which generation dependence of the scalar masses $m_{L_i}^2$, $m_{Q_i}^2$ 
and soft parameters $A_{ikk}^{\l'}$ and $A_{ikk}^{D}$ 
is neglected. Since we are assuming the universal boundary conditions, this
is true in the leading order in which  the $Q^2$ dependence of the
parameters multiplying $\lambda'_{ikk} h_k^D$ in eq.(\ref{rge}) is
neglected.  $Q^2$ dependence in these parameters generated through the 
gauge couplings will also be flavor blind though Yukawa couplings
will lead to some  generation
dependent corrections. But their impact on the 
the conclusions based on the analytic approximation below is not 
expected to be significant. The neglect of the generation dependence of
$\kappa_{ik},\kappa'_{ik}$ allows us to rewrite eq.(\ref{mnot})
as,
\beq
\label{mnotp}
{\cal M}^0_{ij} \equiv m_0 a_i a_j \;\; ,
\eeq
where, 
\beq \label{factor}
a_i \equiv \l'_{ikk}~ h^D_{k}
\eeq
$m_0$ is now completely determined by the standard MSSM parameters and the
dependence of the R-violating parameters gets factored out
 as in eq.(\ref{mloop1}). $m_0$ can be determined by solving the RGE 
(\ref{rge}). Roughly, $m_0$ is given by,
\beq 
\label{expmnot}
m_0 \sim \left( {3 \over 4 \pi^2} \right)^2 {v^2 \over M_{SUSY}} 
\left( ln {M_X^2 \over M_Z^2} \right)^2\;\; .
\eeq
Let us rewrite the loop induced mass matrix as,
\beqa
{\cal M}^l_{ij}&=&m_{loop}~ \l'_{ilk}~ \l'_{ikl}~ h_{k}^D~ 
h_{l}^D \nonumber \\
&=& m_{loop}~ a_i~ a_j + m_{loop}~ h_{2}^D~ h_{3}^D ~ A_{ij} + 
O( h_{1}^D, h_{2}^D)\;\; , 
\eeqa
where, 
\beq
A_{ij} = \l'_{i23} \l'_{j32} + \l'_{i32} \l'_{j23} - \l'_{i22} \l'_{j33} - 
 \l'_{i33} \l'_{j22}\;\; .
\eeq
Neglecting O($h_{1}^D$,$h_{2}^D$) corrections to the loop 
induced mass matrix, the total mass matrix is given by,
\beqa 
\label{mtotal2}
{\cal M}_{ij}^\n&\approx&(m_0 + m_{loop} )~ a_i a_j + m_{loop}~ h_{2}^D~
h_{3}^D~
 A_{ij} \nonumber \\
&\approx& {\cal M}'_{ij} + m_{loop}~ h_{2}^D~ h_{3}^D~ A_{ij} \;\; .
\eeqa
The matrix ${\cal M}'$ has a special structure. It has only one eigenvalue. 
It can be easily diagonalised using a unitary transformation,
\beq
U^T {\cal M'} U = diag~ (0,0,m_3)\;\;,
\eeq
where,
\beqa
\label{mthree}
m_3 &\approx& (m_0 + m_{loop})~ (a_1^2 + a_2^2 + a_3^2) \nonumber \\
&\sim& (m_0 + m_{loop})~~(~\l^{'2}_{333} + \l^{'2}_{233} +
 \l^{'2}_{133} )~~ {h_{3}^D}^2.
\eeqa
The matrix U is determined as,
\beq
U = \left(
\ba{ccc}
c_2&s_2 c_3&s_2 s_3\\
-s_2&c_2 c_3&c_2 s_3\\
0&-s_3&c_3\\ \ea \right)\;\; ,
\eeq
with,
\beq \ba{cc}
s_2={a_1 \over \sqrt{a_1^2 + a_2^2}}~,\;\;\;&s_3={(a_1^2 + a_2^2)
^{1 \over 2} \over \sqrt{a_1^2 + a_2^2 + a_3^2}}\;\; . 
\ea \eeq
The total mass matrix is now given by,
\beqa
U^T {\cal M}^{\n} U &\approx& m_3 ~diag~(0,0,1) + m_{loop}~ h_{2}^D~
 h_{3}^D~ A'\nonumber \\
& \approx& m_3 \left (
\ba{ccc}
\e A'_{11}&\e A'_{12} & \e A'_{13}\\
\e A'_{12}&\e A'_{22} & \e A'_{23}\\
\e A'_{13}&\e A'_{23} &  1  \ea \right )\;\; ,
\eeqa
where,
\beq
A' = U^T A~ U
\eeq
and
$$\e~ A'_{ij} \approx {m_{loop} \over m_3 }~h_2^D~ h_3^D~ A'_{ij}
\approx {m_{loop} \over m_0 }~{h_2^D \over h_3^D}\;\; .$$ 
The last equality follows under the assumption that $\l'_{ijk}$ are
similar in magnitude and $m_{loop} \ll m_0$.

Choosing, 
\beq
U' = \left(
\ba{ccc}
c_1& s_1 &0\\
-s_1&c_1 &0\\
0&0&1 \ea \right)
\eeq
with, $s_1$, $c_1$ defined by,
\beq
\tan 2 \th_1 = { 2 A'_{12} \over A'_{22} - A'_{11}}\;\;,
\eeq
we have,
\beqa
U'^T U^T {\cal M}^{\n} U U'&= &m_3 \left(  \ba{ccc}
\e \d_1&0&c_1 \e A'_{13} - s_1 \e A'_{23}\\
0& \e \d_2&s_1 \e A'_{13} - c_1 \e A'_{23}\\
c_1 \e A'_{13} - s_1 \e A'_{23}&s_1 \e A'_{13} - c_1 \e A'_{23}&1 \ea
\right)\;\;' 
\nonumber \\
&\approx& m_3~diag( \e \d_1, \e \d_2, 1)\;\;.
\eeqa
The off-diagonal elements will generate additional mixing in the model. But,
as $\e A' \ll 1$, we can neglect these off-diagonal elements. The eigenvalues
in this approximation are given as,
\beq \label{evalues}
m_{\n_1}\sim \e~ m_3~\d_1 \;\;\; ; \;\;\; 
m_{\n_2}\sim \e~ m_3~\d_2 \;\;\; ; \;\;\; 
m_{\n_3}\sim m_3\;\; ,
\eeq
where,
\beqa \label{deltas}
\d_1 &= &  ~(c_1^2 ~A'_{11} - 2 c_1 s_1 A'_{12} +
 s_1^2 A'_{22}) \;\;,\nonumber \\
\d_2 &= &   ~(s_1^2 ~A'_{11} + 2 c_1 s_1 A'_{12} +
 c_1^2 A'_{22})\;\; . 
\eeqa

Note that both $\d_1$ and $\d_2$ are generically of O($\l^{'2}$) when all
$\l'_{ijk}$ are assumed to be similar in magnitude. As a consequence, the 
neutrino masses follow the hierarchy,
$$m_{\n_1} \sim m_{\n_2} \ll m_{\n_3}$$
With,
\beq \label{ratio}
{m_{\nu_2}\over m_{\nu_3}}\sim {m_s\over m_b}{m_{loop}\over
m_0}\left({\delta_2\over \Sigma_i \lambda_{i33}^{'2 }}\right)
\eeq
The last factor in the above is of O(1) and the remaining part is controlled
completely by the standard parameters of the MSSM.

Eq.(\ref{ratio}) may be regarded as a generic prediction of the model.
It is seen from eqs. (\ref{expmloop},\ref{expmnot}) that typically,
\beq
{m_{loop} \over m_0} \sim {\pi^2 \over 3 \left( ln~ 
{M_X^2 \over M_Z^2} \right)^2 } \sim  10^{-3}
\eeq
Thus the neutrino mass ratio in eq.(\ref{ratio}) is suppressed considerably
compared to eq.(\ref{heir}) obtained when sneutrino vev contribution is 
completely neglected. The exact value of this suppression factor is dependent
on MSSM parameters and we will calculate it in the next section.

The mixing among neutrinos is governed by,
\beqa \label{mix}
K&=&U~U'\nonumber \\
&=&\left( \ba{ccc}
c_1 c_2 - s_1 s_2 c_3&  s_1 c_2 + c_1 s_2 c_3 & s_2 s_3 \\
-s_2 c_1 - s_1 c_2 c_3&- s_1 s_2 + c_1 c_2 c_3 & c_2 s_3 \\
s_1 s_3 & -s_3 c_1 & c_3 \ea \right)
\eeqa
The angles are determined by the ratios of the trilinear couplings and 
hence can be naturally large. Thus, as in supersymmetric model with
purely bilinear $R$ violation \cite{gauge} one gets hierarchical masses
 and large mixing without fine tuning in any parameters.\\
{\bf 4. Neutrino Anomalies}\\
We now discuss the phenomenological implications of neutrino masses,
eq.(\ref{evalues}) and mixing, eq.(\ref{mix}). Due to hierarchy in 
masses, one could simultaneously solve the solar and atmospheric $\n$
problems provided, $m_{\n_1} \sim m_{\n_2} \sim 10^{-5}$ eV and 
$m_{\n_3} \sim 10^{-2}$ eV. 

In order to determine these masses exactly, we have numerically integrated
eqs.(\ref{rge}) along with similar equations for the parameters 
appearing in them.
We have imposed the standard universal boundary condition and required 
radiative breaking of the $SU(2) \times U(1)$ symmetry. Solution of the RGE
determines both $m_{loop}$ (eq.(\ref{mloop1})) and
 $m_0$ (eq.(\ref{mnotp})). We display these in figs. (1a,1b) as a function of
$\m$ for $tan \b = 2.1 $, $M_2 = 200 $ and $400 \GeV$ respectively. The
ratio ${m_{loop} \over
m_0}$ is
quite sensitive to the sign of $\m$. For $\m >0$, this ratio is rather
small, typically,  
$\sim 10^{-2} - 10^{-3}$, while it can be much larger for
 $\m <0$. There exists a region with negative  $\m$ in which ${m_{loop} \over
m_0}\geq 1$. In this region, two contributions to the sneutrino vev in eq.
(\ref{omega}) cancel and $m_0$ gets suppressed. Barring this region, the
${m_{loop} \over m_0}$ is seen to be around $\sim 10^{-1}-10^{-2}$
 for negative $\mu$ leading to
\beq
{m_{\n_2} \over m_{\n_3}} \sim {m_s\over m_b} {m_{loop}\over m_0}\sim 
2~(10^{-3}-10^{-4})\eeq
For $m_{\n_3} \sim 10^{-1} - 10^{-2} \eV$, one thus
obtains $m_{\n_2} \sim m_{\n_1} \sim 2~(10^{-4} - 10^{-6}) \eV$ which is
in the right range required  to solve the solar neutrino problem through
vacuum oscillations.
The typical value of $m_0 \sim \GeV$ found in Fig.(1b) implies through
eq.(\ref{mthree}),
$$\l' \sim 10^{-4}$$
Thus one needs to choose all $\l'_{ijk}$ of this order. Once this is done,
one automatically obtains solar neutrino scale for some range in the MSSM
parameters.

While, hierarchy needed for the vacuum solution follows more naturally,
one could also obtain scales relevant to the MSW conversion. This happens
for very specific region of parameters with negative $\mu$ in which two
contributions to sneutrino vev, eq. (\ref{omega}), cancel. As already
mentioned,
${m_{loop} \over m_0}$ can be 1 in this region. 
One  then recovers the result of \cite{drees}, namely,
 eq.(\ref{heir}) which allows
MSW solution for the solar neutrino problem. The reference
\cite{chun} which  used
hierarchical $\lambda'_{ijk}$ also concentrated on this region in order to
obtain the MSW solution. 

We showed in Fig. 1 neutrino mass ratio for specific value of
 $M_2$ and tan$\b$.
Qualitatively similar results follow for other values of these parameters.
We have displayed in Table.1 values for the MSSM parameters 
and what they imply for ${m_{\n_2} \over m_{\n_3}}$. We have shown
illustrative values of the parameters which lead to the vacuum as well as
MSW solution. The latter arise only for limited parameter range  
corresponding to
cancellations in eq.(\ref{omega}). The former is a more generic
possibility
which arise for larger region with both positive and negative values of
$\mu$.
The MSW solution in the present context will have to be restricted
to the large angle solution if one does not want to impose any discrete
symmetries or fine tune $\l'$s.  

The constraints on mixing matrix K, eq.(\ref{mix}), implied by the
experimental results are also easy to satisfy keeping all the $\l'_{ijk}$
similar in magnitude. Hierarchy in masses, $m_{\n_2}, m_{\n_3}$ lead to 
the following survival probabilities for the solar and atmospheric neutrinos
after undergoing vacuum oscillations:
\beqa
P_e&=& 1 - 4 ~K_{e1}^2~ K_{e2}^2~ \sin^2 \left({ \Delta m_{21}^2 t \over 4 E }
\right)  - 2 ~K_{e3}^2 ~(1 - K_{e3}^2) \\ \nonumber
P_\m &=& 1 - 4 ~K_{\m_3}^2(1- K_{\m_3}^2) \sin^2 \left
({ \Delta m_{31}^2 t \over 4 E } \right)\;\;.
\eeqa
Where, $\Delta m^2_{ij} = m_{\n_i}^2 - m_{\n_j}^2$ and 
$\Delta m^2_{31} \sim \Delta m^2_{32}$. 

These survival probabilities assume the standard two generation form when
$K_{e3} = 0 $ and 
one could utilize existing constraints on mixing angles.
In practice, the $K_{e3}$ may not be zero but is constrained to be 
small from the non-observation \cite{choose}
 of $\n_e$ oscillations at CHOOZ.  To the 
extent it is small,  one could use the two generation constraints for the
solar and atmospheric analysis. The combined constraints which are needed
\cite{sk,bsk} to be satisfied are:
\beqa
0.6 ~\leq~ 4~ K_{\m 3}^2~ (1 - K_{\m 3}^2) & = & s_3^4~ \sin^2 2 \theta_2 
~+ ~ c_2^2~ \sin^2 2 \theta_3~ \leq ~1. \nonumber \\
K_{e3} &~\leq~& 0.18 \\ \nonumber
0.8 ~\leq~ 4~ K_{e1}^2~ K_{e2}^2 &=& 4 (c_1 c_2 - s_1 s_2 c_3)^2~~
 (s_1 c_2 + s_2 c_1 c_3 )^2 ~\leq~ 1.
\eeqa
It is possible to satisfy all these constraints by choosing for example,
$$c_3 = s_3 = s_1 = c_1 = {1 \over \sqrt 2 }~;~~ s_2 = 0.28 $$
The relative smallness of $s_2$ required here does not imply significant
fine tuning and can be easily obtained, e.g. by choosing,
$${\l'_{133} \over \l'_{233}} \sim {1 \over 3}$$. 

We have so far concentrated on the $\l'_{ijk}$ couplings alone. The analogous
discussion can be carried out for $\l_{ijk}$ couplings appearing in the
eq.(\ref{wl}). Here also, the tree level contribution to neutrino masses will
dominate over the loop contribution although the structure of mixing matrix
will differ slightly due to the anti-symmetry of the couplings $\l_{ijk}$ 
in indices $i$ and $j$.\\ 
{\bf 5. Discussion :}\\
We have discussed in detail the structure of
 neutrino masses and 
mixing in MSSM in the presence of trilinear R-violating couplings, 
specifically
$\l'_{ijk}$. Noteworthy feature of the present analysis is that it is
possible to obtain the required neutrino mass pattern under fairly general
assumption of  all the $\l'_{ijk}$ being of equal magnitudes. This is to
be contrasted with the recent analysis \cite{drees,chun,arc} which had to
make very
specific
choice of the trilinear couplings in order to reproduce neutrino mass
pattern. It is quite interesting that hierarchy
among neutrino masses is controlled by few parameters in MSSM and is
largely independent of the trilinear $R$ violating couplings. Thus
one could understand the required neutrino mass ratio without being 
specific about the exact values of large number of the trilinear
couplings. This `model-independence' is an attractive feature of the
scenario discussed here.

The key difference of the present work compared to many of the other
works is  proper inclusion of the sneutrino vev
contribution. While we had to resort to specific
case of the minimal supergravity model for calculational purpose, the
sneutrino vev
contribution would arise in any other scheme with $\l'_{ikk} \neq 0$ at 
a high scale such as $M_{GUT}$. Such contribution thus cannot be
neglected a priori. On the contrary, the inclusion of this
contribution makes the model more
interesting and fairly predictive in spite of the presence of large number
of unknown couplings.

\newpage
\textwidth 20.0truecm
\begin{center}
\begin{tabular}{|c|c|c|c|c|c|}
\hline
~~~~~ m~~~~~  &$~~~~~ M_2~~~~~  $&~~~~~ $\m$~~~~~  &~~~~~ $m_0$~~~~~ 
& ~~~~~$m_{loop}~~~~~$ 
&~~~~~$ratio$~~~~~ \\
 ~~~~~GeV~~~~~ & ~~~~~GeV~~~~~ & ~~~~~GeV~~~~~ & ~~~~~GeV~~~~~ 
& ~~~~~GeV~~~~~ & ~~~~~${m_{\n_2}
\over
m_{\n_3}}$~~~~~ \\
\hline
 1312 &200&1225&-16.94&-0.1689&$1.8~10^{-4}$\\
 1056 &250&1100&-23.07&-0.2010&$1.6~10^{-4}$\\
 3898 &300&-3400&-3.238&0.0394&$2.3~10^{-4}$\\
 3921 &350&-3450&-2.655&0.0376&$2.6~10^{-4}$\\
\hline
 225.7 &400&-1038&-0.0368&0.0369&$-1.8~ 10^{-2}$\\
 192.9 &350&-907&-0.0423&0.0420&$-1.8~~10^{-2}$\\
 157.2 &300&-777.5&-0.0470&0.0487&$-1.9~~10^{-2}$\\
\hline
\end{tabular}
\end{center}
\vskip 1.5truecm
{\bf Table1:} {\sl The values of $m_0$,  
$m_{loop}$ and ratio of the eigenvalues, ${m_{\n_2} \over m_{\n_3}}$
  for various values of the standard MSSM parameters m, $M_2$ 
and $\m$ for tan $\beta = 2.1$ , $A$=0. }

\newpage
\begin{figure}[h]
\epsfxsize 15 cm
\epsfysize 15 cm
\epsfbox[25 151 585 704]{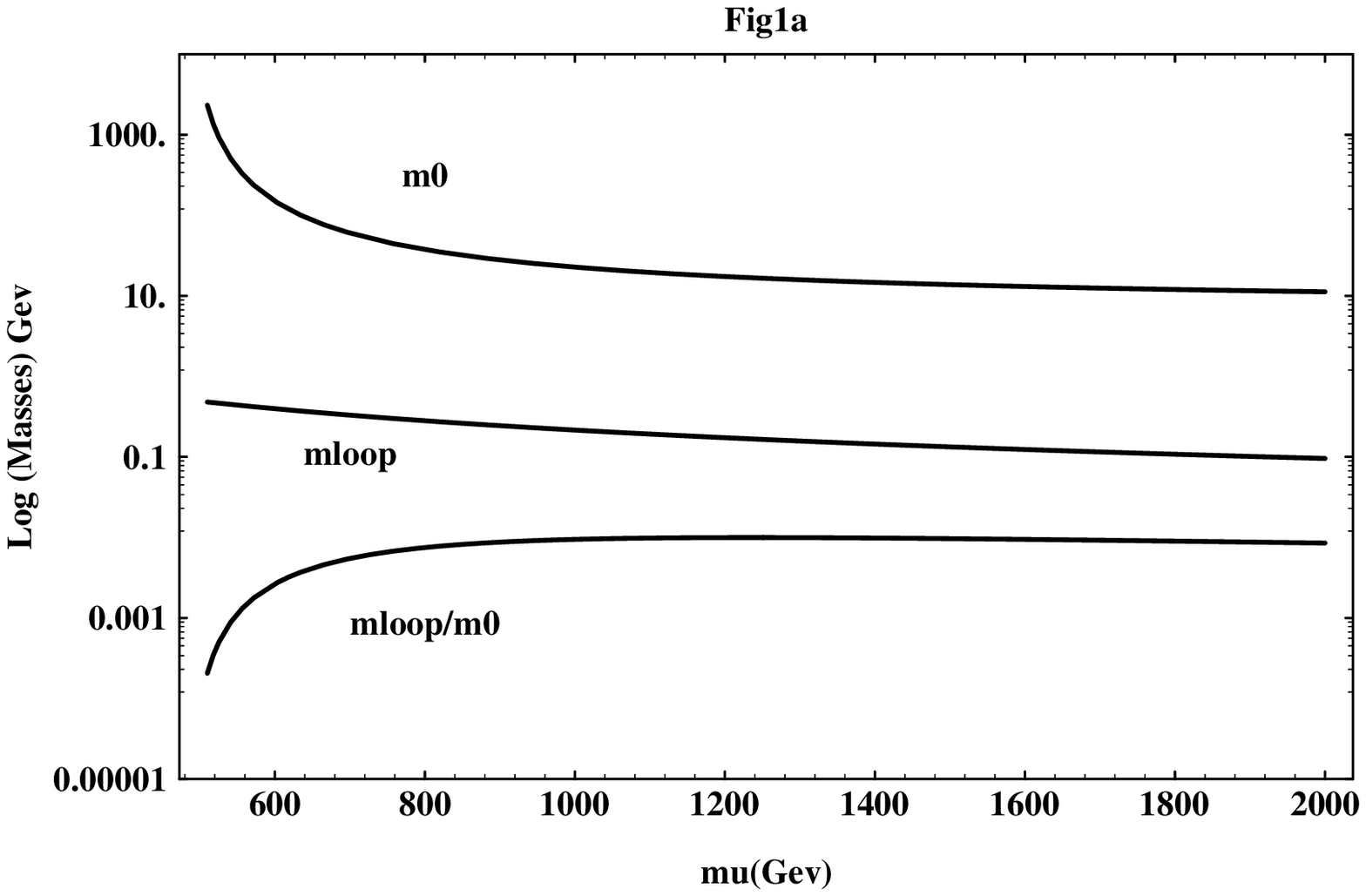}
\label{fig1a}
\end{figure}
\vskip 1.0truecm
\noindent
{\bf Figure 1a}. {\sl  The absolute values of tree level contribution,
$m_0$, the loop level
contribution, $m_{loop}$  and  their ratio ${m_{loop} 
\over m_0}$ are plotted with respect to $\m$ (positive) for 
$M_2 = 200\;GeV$, $A$=0 and tan$\beta$ = 2.1 . The $m_0$ and
$m_{loop}$
are defined in the text.
\newpage
\begin{figure}[h]
\epsfxsize 15 cm
\epsfysize 15 cm
\epsfbox[25 151 585 704]{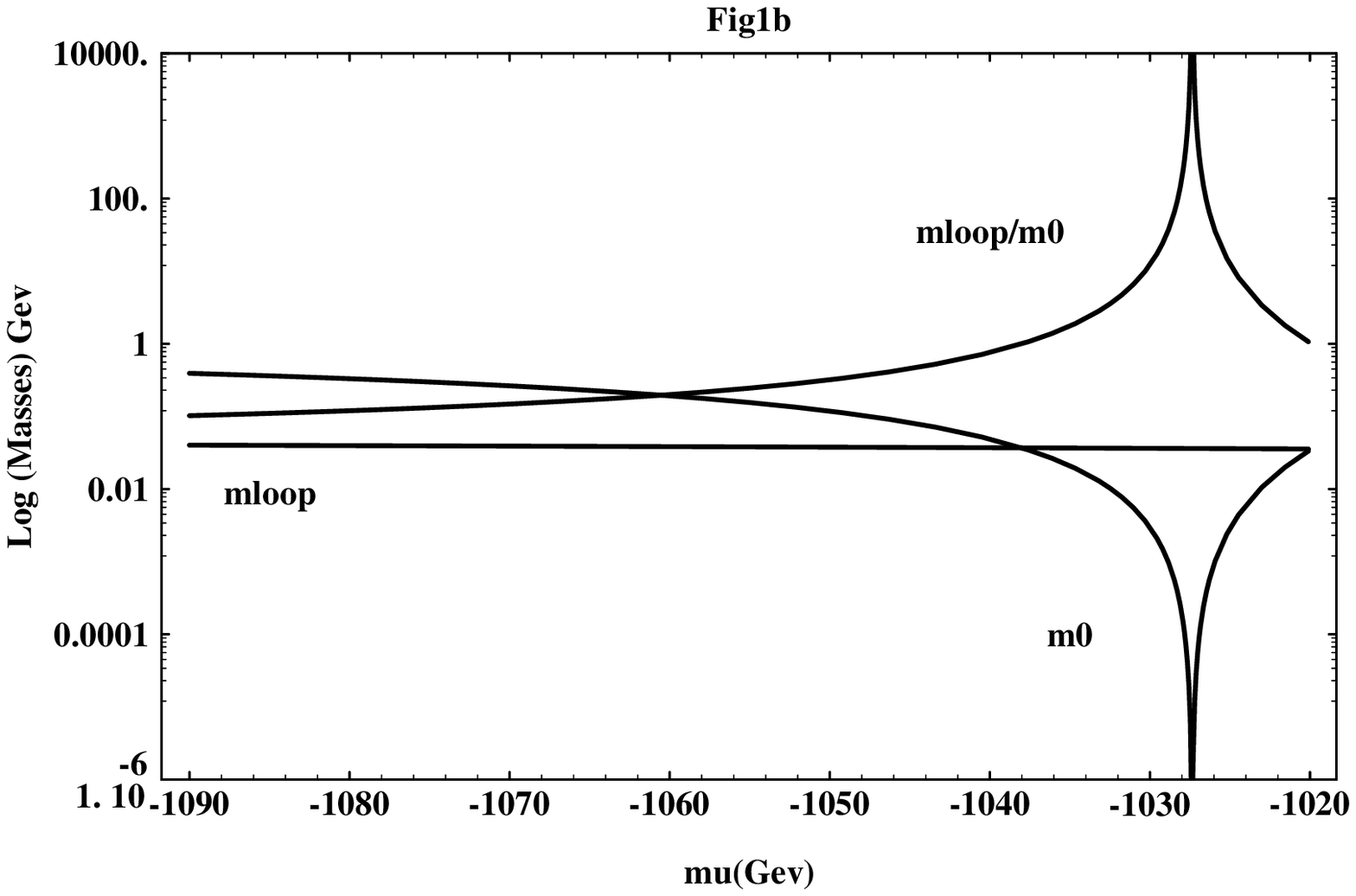}
\label{fig1b}
\end{figure}
\vskip 1.5truecm
\noindent
{\bf Figure 1b}. {\sl  Same as in Fig. (1a) but  for $M_2 = 400 \;GeV $
and
$\m$ (negative). }

\begin{thebibliography}{99}
\bibitem{f2}It should be noted
that models considered in \cite{drees} contain diagonal $\l'_{ikk}$ even
after imposition of the discrete symmetries and hence contain the tree level
neutrino mass.
\end{thebibliography}
\end{document}